\documentclass{osa-article}

\journal{oe}

\usepackage[T1]{fontenc}
\usepackage[english]{babel}

\usepackage{amsmath}
\usepackage{bm}
\usepackage{enumitem} 

\usepackage[normalem]{ulem}

\usepackage{textcomp}
\usepackage{subfigure}

\begin{document}

\title{Raman scattering model of the spin noise}

\author{G.~G.~Kozlov,\authormark{1} A.~A.~Fomin\authormark{1}, M.~Yu.~Petrov\authormark{1,*}, I.~I.~Ryzhov\authormark{2,1} and V.~S.~Zapasskii\authormark{1}}

\address{\authormark{1}Spin Optics Laboratory, Saint~Petersburg State University, St.~Peterbsurg, 198504 Russia \\
\authormark{2}Department of Photonics, Saint~Petersburg State University, St.~Peterbsurg, 198504 Russia}

\email{\authormark{*}m.petrov@spbu.ru} 



\begin{abstract}
	The mechanism of formation of the polarimetric signal observed in the spin noise spectroscopy (SNS) is analyzed from the viewpoint of the light scattering theory. A rigorous calculation of the polarimetric signal (Faraday rotation or ellipticity) recorded in the SNS is presented in the approximation of single scattering. We show that it is most correctly to consider this noise as a result of scattering of the probe light beam by fluctuating susceptibility of the medium. Fluctuations of the gyrotropic (antisymmetric) part of the susceptibility tensor lead to appearance of the
 typical for the SNS Faraday rotation noise at the Larmor frequency. At the same time, fluctuations of linear anisotropy of the medium (symmetric part of the susceptibility tensor) give rise to the ellipticity noise of the probe beam spectrally localized at the double Larmor frequency. The results of the theoretical analysis well agree with the experimental data on the ellipticity noise in cesium vapor. 
\end{abstract}

\section{Introduction}
Spin noise spectroscopy (SNS), for the first time demonstrated on a sodium vapor~\cite{zap-frns1981} back in 1981, has been rapidly developing over the past decade and has proven to be an efficient tool for studying the energy structure of matter (see, e. g., reviews \cite {muller-semicon-sns2010,zapasskii-sns-review2013,glazov-linear-optics-raman-sns2015,sinitsyn-theory-sns-review2016,smirnov-sn-review2020}).
To a large extent, the success of the SNS was inspired by its applicability for studies of semiconductor systems~\cite{oestreich-sns-gaas2005,romer-sns-review2007,crooker-sn-gaas2009,huebner-rapid-scanning-of-sn2013,gaas-high-temp-kuhn2017}, including nanostructures~\cite{muller-mqw-sns2008,crooker-sns-e-holes-qd2010,wiegand-reoccup-noise-sqd2018}. Not only spin carrier Larmor precession in volume (\cite{crooker-sn-gaas2009,kamenskii-re-sns2020} etc.) and low-dimensional media (e.\,g.~\cite{poltavtsev-sns-single-qw2014,smirnov-noneq-sn-qd2017}) can be addressed by SNS, but also the nuclear dynamics~\cite{smirnov-cooled-nucleui-sn2015,ryzhov-nuclear-dynamics-sns2015,vladimirova-spin-temp-concept2018}, light-matter interaction effects~\cite{ryzhov-sn-explores-local-fields2016,poltavtsev-gyrotropy-asymmetric2014}, spatial properties distribution~\cite{yang-two-color-sns2014,cronenberger-spatiotemporal-sns2019}, and even noise of valley redistribution of electrons~\cite{goryca-valley-noise2019}. Still the atomic systems attract the significant attention of researchers~\cite{lucivero-corr-func-sn-diffusion2017,ma-osns-rb-gas-hom-inhom2017,petrov-homogenization-doppler2018,swar-meas-spin-prop-atomic2018,fomin-spin-align-noise2020,tang-sns-hot-vapor-antirelax2020,vershovskii-projection-sn2020,zhang-sns-pulse-modulated2020} as model objects which allow to reveal the fundamental pecularities of spin noise formation mechanics.
In typical SNS experiments with polarimetric detection of the signal, one measures, in fact, the magnetization-noise power spectrum of the studied sample associated, in accordance with the fluctuation-dissipation theorem, with the spectrum of its magnetic susceptibility~\cite{kubo-fdt1966}. 
The experiments are usually carried out in a transverse static magnetic field, so that the magnetization-noise power spectrum turns out to be localized in the region of the Larmor frequency $\omega_L$of precession of the magnetic moments (spins) and represents, in fact, EPR (electron paramagnetic resonance) spectrum of the system.
The fact that magnetization noise is observed in SNS as the polarization noise of the probe laser light and is caused by fluctuations of the optical susceptibility of the sample shows a similarity of the SNS and Raman spectroscopy.
Thus, due to combination of the features of the EPR and Raman spectroscopy, the SNS reveals a number of unique features related to nonperturbativity of the SNS~\cite{cronenberger-atomic-like-sn2015,scalbert-fundamental-limits2019}, to possibility of tuning the frequency of the probe laser light~\cite{zapasskii-optical-sns2013,dahbashi-sn-single-hole2014} and its spatial localization~\cite{kozlov-heterodyne-detect-scatt-tomography2018}, to possibility of using multi-beam configurations~\cite{pershin-two-beam-sns2013,kamenskii-amplif-scatt-sn-qd2020} and some others.

Interpretation of the SNS experiments is performed, as a rule, using ``magnetic'' language with the observed Faraday rotation (FR) noise being identified with the magnetization noise and the FR power spectrum calculated using appropriate model.
Meanwhile, strictly speaking, the fluctuations of the scattered probe light observed in the SNS experiments, can be caused not only by fluctuations of gyrotropy (proportional to the magnetization), but also by fluctuations of the whole optical susceptibility tensor of the medium~\cite{ryzhov-sns-amplification2015}. 

The idea of interpretation of the spin noise signal formation mechanism in terms of Raman scattering is well known to the audience involved in the SNS and was suggested shortly after the first experiment of Aleksandrov and Zapasskii~\cite{gorbovitskii-aleksandrov-zapasskii1983}. The aim of our work work is to rigorously develop this idea and to fill the existing theoretical gap with a consistent description of the mechanism of formation of the light-polarization noise observed in SNS as a noise of light scattering.

The work is organized as follows. 
In the first section, we present a brief description of a typical SNS setup and derive the general relationship between the polarimetric signal (FR or ellipticity noise) recorded in SNS and optical susceptibility tensor of the scattering particles. 
In the second section, we calculate, in the framework of a simple atomic model, the optical susceptibility tensor for the case of nonstationary (superposition) atomic state.
The derived expression is used in the third section for calculating
the SNS polarimetric signal. In this section, we analyze spectral peculiarities of the noise signal and their dependence on the angle between the direction of the probe beam polarization plane and magnetic field. 
In the fourth section, we analyze, in the framework of the developed model, the mechanism of formation of the recently discovered ellipticity noise signal at double Larmor frequency (alignment noise) and compare our results with the experimental data obtained in \cite{fomin-spin-align-noise2020}.
 The main results of the paper are summarised in Conclusions.

{Throughout the article, polarimetric signals recorded in SNS are considered as the result of Raman scattering in its simplest form, when the modulation of the optical susceptibility leads to the appearance of side frequencies in the scattered field, which is described by classical electrodynamics. This assumption seems to be justified for the purposes of SNS, since the frequency shift of the scattered radiation is much less than the temperature (in frequency units) at which the measurements are made. }
\section{The relationship between SNS polarimetric signal and susceptibility tensor}

\begin{figure}
 	\begin{center}
 		\includegraphics[width=0.8\linewidth]{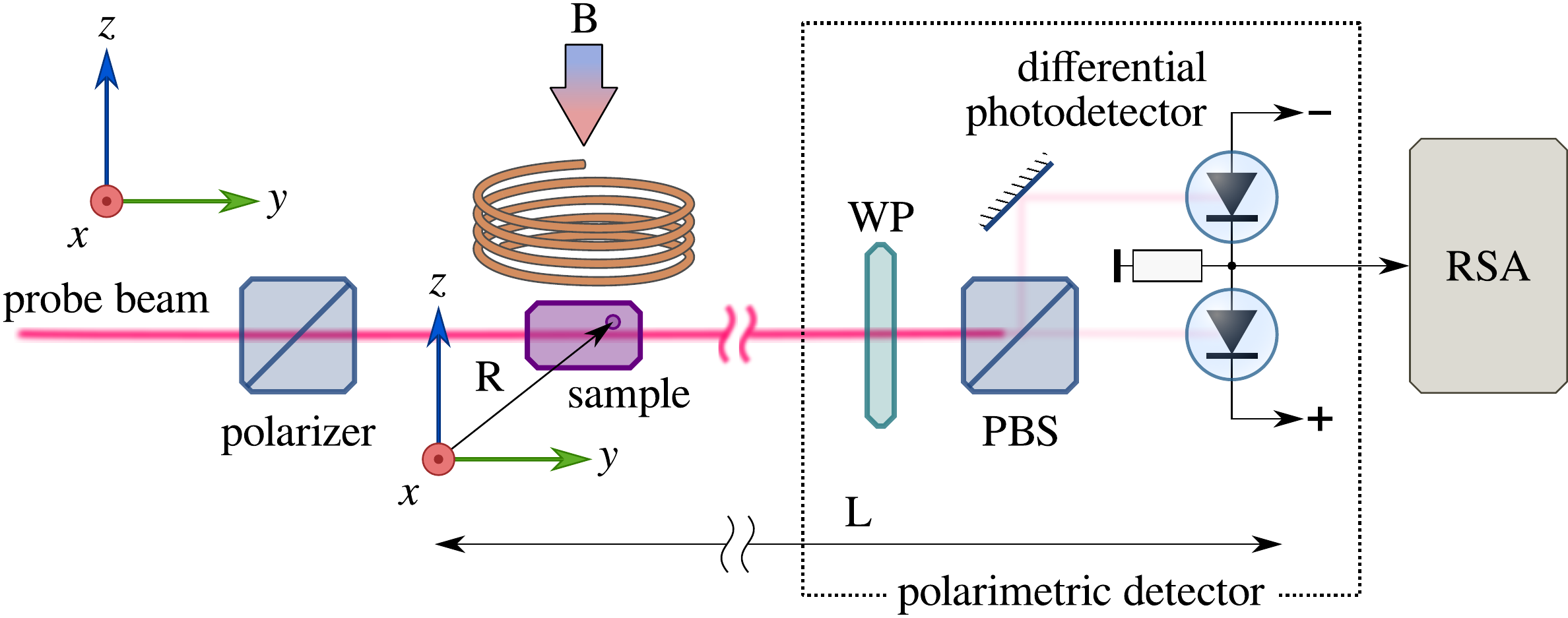}
 		\caption{Schematic of the experimental setup. When observing fluctuations of the Faraday rotation, a $\lambda/2$ waveplate (WP) was placed and used to balance photocurrents of the detector. When observing fluctuations of ellipticity, a $ \lambda / 4 $ WP was placed with its axes aligned at 45$^\circ$ with respect to the PBS eigen directions. }
 		\label{fig1}
 	\end{center}
 \end{figure}

 Schematic of the experimental setup used for the SNS measurements is shown in Fig.~\ref{fig1}. 
 The polarimetric detector in this setup could detect either FR or ellipticity noise depending on the type of the waveplate (WP) placed in front of the polarization beamsplitter (PBS). When detecting the FR noise, the WP was taken half-wave and was used to balance the differential photodetector. When detecting the ellipticity noise, the WP was taken quarter-wave, with its axes aligned at 45$^\circ$ with respect to axes of the beamsplitter PBS. 
In this arrangement of the polarization elements, the output signal of the differential photodetector was zero, when the input light was polarized linearly (regardless of the polarization plane azimuth) and became nonzero only upon appearance of ellipticity in the input light. 
The output electric signal $U$ of the polarimetric detector was fed to a digital hardware Fourier-transform spectrum analyzer which acquired the polarization noise spectrum of the probe beam transmitted through the sample. 
Magnetic field could be aligned either along or across the probe-light propagation axis (in the Faraday or Voigt geometries).

 As was shown in~\cite{gorbovitskii-aleksandrov-zapasskii1983}, the polarization noise signal observed in SNS can be interpreted as the result of scattering of the probe beam by the studied sample comprising any paramagnetic particles (atoms, ions, electrons or either quasiparticles such as holes and excitons), in what follows they will be referred to as {\it atoms}.
 We will calculate this signal using the following simplifying assumptions:
 \begin{enumerate}[label=(\roman*)]
\item The electromagnetic field acting upon each atom coincides, with an acceptable accuracy, with that of the probe beam (approximation of single scattering);
\item Atomic polarization can be calculated in the approximation of linear response; 
\item The magnetic field is small enough to Zeeman splitting of the atomic multiplets $\sim\omega_L$ be much smaller than the homogeneous linewidth $\delta$ of the optical transition and not be resolved in the optical spectrum ($\omega_L \ll \delta$).
\end{enumerate}

\subsection{Polarimetric response of a single atom}

In this subsection we consider the case of {\it ellipticity} noise -- the FR noise spectrum can be calculated in a similar way ~\cite{kozlov-light-scat-appl-sns2017}). 
In this case, the quarter-wave plate of the polarimetric detector (Fig.~\ref{fig1}) is aligned with its axes at $45^\circ$ with respect to polarization directions of the polarizing beamsplitter (PBS), which we assign to be the axes $z$ and $x$. 
The probe beam propagation direction is taken for the axis $y$ (see Fig.~\ref{fig1}). 
In this coordinate system, the magnetic field has only $z$~component (in Voigt geometry), ${\bf B}=(0,0,B)$, while the probe-beam electric field $\bf {\cal E}_0$ has only the $x$ and $z$~components, ${\bf {\cal E}}_0=({\cal E}_{x0},0,{\cal E}_{z0})$ (Fig.~\ref{fig1}).

Let us denote electric field of the probe beam at the input of the polarimetric detector as ${\bf {\cal E}}=({\cal E}_{x},{\cal E}_{y},{\cal E}_{z})$. 
Then, as can be shown by direct calculations, the output signal $U$ of the detector, in this mode of operation, is given by 
\begin{equation}
U={\frac{1}{T}}\int_0^Tdt \int_S dx dz \bigg [ {\cal E}_x(t){\cal E}_z(t+\Delta) - {\cal E}_x(t+\Delta) {\cal E}_z(t)\bigg ], 
\label{1}
\end{equation} 
where $\Delta={\pi/ 2\omega}$ and $\omega$ is the probe light frequency.
The integration over $dxdz$ in Eq.~\eqref{1} is performed over photosensitive surfaces of the photodetectors $S$, which are supposed to be identical. 
The integration over $t$ corresponds to averaging over the time interval $T$ that contains integer number of optical periods and meets the requirement $2\pi/\omega \ll T \ll 2\pi/ \omega_L$. 
We see that, indeed, the signal $U$ appears to be nonzero only for the elliptically polarized input field ${\bf {\cal E}}_0$, while for any linearly polarized field, the output signal vanishes. 
It is also seen from Eq.~\eqref{1} that any rotation of the polarimetric detector around the $y$~axis does not affect the output signal $U$, since, for any two vectors $\bf A$ and $\bf B$, the quantity 
\begin{equation}
A_xB_z-A_zB_x= ({\bf A,\beta B}),\hskip3mm\beta\equiv \begin{pmatrix} 0 & 1 \\ -1 & 0 \end{pmatrix} 
\label{metric}
\end{equation}
does not change under arbitrary rotations in the plane $xz$. 

The noise signal observed in SNS experiments can be represented as the sum of the contributions of individual atoms. 
Therefore, we will start with calculating the ellipticity signal $\delta u_e$, created by a single atom. 
For convenience, we will consider the field $\cal E$ at the input of our detector as a real part of the complex field ${\bf E}$: ${\cal E}= \mathop{\rm Re} {\bf E}$. 
The field ${\bf E}$ can be considered as a sum of the field ${\bf E}_0\equiv {\bf A}_0 \exp{-\imath\omega t}$ of the probe beam $({\cal E}_0= \mathop{\rm Re} {\bf E}_0)$ and the field ${\bf E}_1$ created by the atomic dipole $({\cal E}_1= \mathop{\rm Re} {\bf E}_1)$. 
We use calligraphic letters to denote the observed (real) fields.

We calculate the ellipticity signal in the approximation linear in the field ${\bf E}_1$. 
Since all the fields are assumed quasi-monochromatic ($\sim \exp{-\imath\omega t}$), the time shift by $\pm \pi/2\omega$ is equivalent to multiplication by $\mp \imath$. 
Keeping this in mind, we obtain from Eq.~\eqref{1} the following expression for the signal $\delta u_e$\ \cite{for}
\begin{equation}
\delta u_e= \hbox{ Im }
{\frac{2}{T}}\int_0^Tdt \int_S dx dz \bigg [ {\cal E}_{0x}{ E}_{1z} -{E}_{1x} {\cal E}_{0z} \bigg ].
\label{2}
\end{equation}
This formula includes both complex $(E_{1i})$ and real $({\cal E}_{0i})$ fields.
The field ${\bf E}_1$ of the atomic dipole can be obtained by solving the inhomogeneous Helmholtz equation $\Delta {\bf E}_1 +k^2{\bf E}_1=-4\pi k^2{\bf P}$, where $k=\omega /c$ ($c$ is the speed of light) and $\bf P(r)\sim \delta ({\bf R-r})$ is the complex polarization created by the atom with the radius-vector $\bf R$. 
Solution of this equation can be obtained using Green's function of the Helmholtz operator (see, e.\,g.,~\cite{kozlov-light-scat-appl-sns2017,kozlov-heterodyne-detect-scatt-tomography2018}) and has the form ${\bf E}_1({\bf r})=k^2\int d^3{\bf r'}\exp{\imath k|{\bf r-r'}|}{\bf P(r')}/|{\bf r-r'}|$. 
By substituting this expression into Eq.~\eqref{2}, we have:

\begin{equation}
\delta u_e={2k^2\over T} \hbox{ Im } \int_0^Tdt \int d^3{\bf r'} \bigg [
\Phi_x({\bf r'})P_z({\bf r'})-\Phi_z({\bf r'})P_x({\bf r'})
\bigg ],
\label{3}
\end{equation}
where we introduced the following functions $\Phi_i({\bf r'}) (i=x,z)$\cite{f0}:
\begin{equation}
\begin{gathered}
\Phi_i({\bf r'})\equiv \int _S dxdz \hskip1mm {\cal E}_{0i}(x,y,z){\exp{\imath k |{\bf r-r'}|}\over |{\bf r-r'}|}\bigg |_{r_y=L} \equiv\\ 
\exp({-\imath\omega t})\Phi_i^+({\bf r'})+\exp({\imath\omega t})\Phi_i^-({\bf r'}), \hskip3mm i=x,z,
\hskip3mm {\bf r}=(x,y,z).
\label{4}
\end{gathered}
\end{equation}

When integrating over the photodetector surface $S$ ($dxdz$) in Eq.~\eqref{4}, we assumed that $r_y=y=L$, where $L$ is the distance from the atom to the polarimetric detector, which we consider to be large: $L\rightarrow \infty$. 
Besides, in Eq.~\eqref{4} we separate explicitly the components $\Phi_i^\pm({\bf r'})$ of the function $\Phi_i({\bf r'})$ proportional to $\exp({\mp\imath\omega t})$. 
In the approximation of linear response, atomic polarization is proportional to the probe wave electric field. 
Therefore, ${\bf P(r')}\sim \exp({-\imath\omega t})$ and, after time-averaging in Eq.~\eqref{3}, only components $\Phi_i^-({\bf r'})$ survive. 
We show in Appendix that (the similar result for the case of Gaussian beam has been obtained in~\cite{kozlov-heterodyne-detect-scatt-tomography2018})
\begin{equation}
\Phi_i^-({\bf r'}) ={\imath \pi\over k} A_{0i}^\ast({\bf r'}), \quad
|{\bf r'}|\ll L, \quad i=x,z;
\label{5}
\end{equation} 
where $A_{0i}$ is the $i$-th projection of the amplitude of the probe beam complex field (remind that ${\bf E}_0=(A_{0x},0,A_{0z})\exp({-\imath\omega t})$ ).
The polarization ${\bf P(r')}$ created by a single atom entering Eq.~\eqref{3} can be presented in the form $P_i({\bf r'})=\delta({\bf r'-R})\langle d_i\rangle \exp({-\imath\omega t})$ where $\langle d_i\rangle $ is the complex amplitude of oscillation of the $i$-th component of the atomic dipole moment, $\bf R$ is the radius-vector of the atom. 
By substituting this expression into Eq.~\eqref{3} and taking into account Eq.~\eqref{5}, we obtain, for the ellipticity signal $\delta u_e$ created by a single atom, the following expression:
\begin{equation}
\delta u_e = 2\pi k\hbox{ Re }\bigg [
A_{0x}^\ast({\bf R})\langle d_z\rangle -A_{0z}^\ast({\bf R})\langle d_x\rangle
\bigg ].
\label{6}
\end{equation} 

When the polarimetric detector operates in the FR-detection mode (i.e., the $\lambda/4$ wave plate is replaced by the $\lambda/2$ plate), then a similar calculation leads to the following expression for the FR noise signal $\delta u_r$ produced by a single atom:
\begin{equation}
	\thinmuskip=-1mu
    \medmuskip=0mu
    \thickmuskip=2mu
\delta u_r=2\pi k \mathop{\rm Im} \bigg [\cos [2\phi]\bigg (A_{0x}^\ast({\bf R})\langle d_x\rangle-A_{0z}^\ast({\bf R})\langle d_z\rangle\bigg ) -\sin[2\phi]\bigg (A_{0x}^\ast({\bf R})\langle d_z\rangle + A_{0z}^\ast({\bf R})\langle d_x\rangle\bigg )\bigg ].
\label{7}
\end{equation} 
Here, $\phi$ is the angle between $z$-axis and one of the main directions of the beamsplitter (in the above calculation of the ellipticity signal, we used a coordinate system for which $ \phi = 0 $; see\ \cite{fbs} for explanation).
 
Equations \eqref{6} and \eqref{7} can be simplified under the following conditions. 
First, we assume that the quantities $\langle d_i\rangle$ entering Eqs.~\eqref{6} and \eqref{7} can be expressed through the probe light electric field amplitude ${\bf A}_0({\bf R})$ with use of the susceptibility tensor $\alpha$: $\langle d_i\rangle =\alpha_{ik}A_{0k}$. 
Second, the probe beam is assumed to be linearly polarized. 
In this case, $A_{0x}=A_0\sin\theta$ and $A_{0z}=A_0\cos\theta$, where $\theta$ is the angle between the probe beam polarization and $z$ axis. 
And third, we assume that, in the measurements of the Faraday-rotation noise, the orientation of the polarimetric detector (specified by the angle $\phi$ or, which is the same, by the orientation of the half-wave plate, see \cite{fbs}) corresponds to conditions of balance with no DC signal at the output of the detector, i.\,e., $\phi=\theta+\pi/4$ \cite{f11}.
When the above conditions are satisfied, Eqs.~\eqref{6} and \eqref{7} can be rewritten in a compact scalar form as follows 
\begin{equation}
\begin{gathered}
\delta u \equiv \delta u_e+\imath \delta u_r=
2\pi k({\bf A}_0({\bf R}),\beta\alpha {\bf A}_0({\bf R}))= \\ 
= \pi k|A_0({\bf R})|^2\bigg [
\alpha_{zx} - \alpha_{xz} - 
(\alpha_{xz}+\alpha_{zx})\cos 2\theta + (\alpha_{zz}-\alpha_{xx})\sin 2\theta
\bigg ]
\label{8}
\end{gathered}
\end{equation}
with matrix $\beta$ defined by Eq. (\ref{metric}).
 
\section{Calculation of the atomic susceptibility}

Calculation of the linear atomic susceptibility is performed assuming that it is related to optical transitions between two (ground and excited) atomic multiplets \cite{f12} with the same total angular momenta $F$. 
Formation of the optical response of the atom travelling through the probe beam can be imagined in the following way. 
The wavefunction of the atom $\Psi(0)$, at the moment of its entering the beam (let it be $t=0$) is a random superposition of atomic eigenfunctions $|1M\rangle$ of the ground multiplet with different $z$-components $M$ of the angular momentum: 
$\Psi(0)=\sum_{M=-F}^FC_M|1M\rangle=\sum_{M=-F}^F|C_M|\exp({\imath \beta_M})|1M\rangle$, where $|C_M|$ and $\beta_M$ are the random amplitude of the atomic state $|1M\rangle$ and its phase (with $\sum_{M=-F}^F|C_M|^2=1$). 
When the magnetic field is nonzero, the ground multiplet exhibits Zeeman splitting, and the above superposition state appears to be nonstationary (even neglecting the probe-beam induced perturbation). 
The appropriate unperturbed density matrix of the atom $\rho_0$ also appears to be time-dependent with nonzero matrix elements only in the subspace of the states of the ground atomic multiplet: 
\begin{equation}
\langle 1M| \rho_0(t)|1M'\rangle=
|C_M||C_{M'}|\exp({\imath [\beta_M-\beta_{M'}]}) \exp({\imath \omega_{1L}[M-M']t}),
\label{9}
\end{equation}
where $\omega_{1L}$ is the Larmor frequency for the ground-state multiplet. 
As seen from Eq.~\eqref{9}, the density matrix oscillates in time at frequencies integer multiples of the Larmor frequency $\omega_{1L}$. 
Since the linear optical susceptibility of the atom is related to its unperturbed density matrix (this connection will be presented below), it may depend on time at frequencies $\omega_{1L}[M-M']$. 
This may, in turn, give rise to appearance of shifted frequencies $\omega\pm \omega_{1L}|M-M'|$ in the spectrum of the field ${\bf E_1}$ scattered by the atom (the effect of Raman scattering) and can be detected in SNS experiments as the noise of polarimetric signal spectrally localized in the vicinity of the frequencies $\omega_{1L}|M-M'|$. 
As shown below, only frequencies with $|M-M'|=0,1,2$ can be observed and, correspondingly, only spectral features at the frequencies $0$, $\omega_{1L}$, and $2\omega_{1L}$ can arise in the polarization noise spectra. 

Let us pass now to calculation of the linear atomic susceptibility. 
Here we present calculations only for the most interesting from the experimental viewpoint case of Voigt geometry as the case of Faraday geometry can be analyzed in a very similar way.
The matrix of the Hamiltonian of the atom in the representation of the two (ground and excited) multiplets in frequency units has the form
\begin{equation}
\begin{gathered}
H=H_0+H_E, \\
H_0 \equiv \Omega \begin{pmatrix} I & 0\\ 0 &0 \end{pmatrix} + \begin{pmatrix} \omega_{2L} J_z & 0\\ 0 & \omega_{1L}J_z \end{pmatrix},\\
H_E= \omega_{x}\exp({-\imath\omega t}) \begin{pmatrix} 0 & J_x \\ J_x & 0 \end{pmatrix}
+\omega_{z}\exp({-\imath\omega t}) \begin{pmatrix} 0 & J_z \\ J_z & 0 \end{pmatrix}
\label{10}
\end{gathered}
\end{equation}
{ Here $H_0$ is the Hamiltonian of undisturbed atom with $\Omega$ being the spectral distance between ground and excited multiplets. The second term in the expression for $H_0$ describe  Zeeman splitting with $\omega_{iL}$ being Larmor frequencies of $i$-th multiplet ($i=1$ refers to the ground state and $i=2$ the excited state).  Term $H_E$ describe the action of  monochromatic $\sim e^{-\imath\omega t}$ electromagnetic field of the probe beam  which is linearly polarised in $ZX$ plain. The quantities $\omega_{x,z}$ correspond to  Rabi frequencies defined by the dipole moment $d$ of  optical transition between the ground and excited multiplets and by the  amplitudes $A_{0x,z}$   of    electric field projections of the probe beam   at point $\bf R$ where the atom is located: $\omega_{i}\equiv dA_{0i}({\bf R})/ \hbar, i=x,z$.} 
Each `element' of matrices in Eq.~\eqref{10} is itself a matrix with dimensions $(2F+1)\times(2F+1)$, with $J_z$ and $J_x$ being known matrices of the corresponding projections of the angular momentum $F$\ \cite{Landau}. 
The matrices of the operators for the needed $x$ and $z$~projections of the atomic dipole moment have the form

\begin{equation}
d_x\equiv d \begin{pmatrix} 0 & J_x \\ J_x & 0 \end{pmatrix}, \quad
d_z\equiv d \begin{pmatrix} 0 & J_z \\ J_z & 0 \end{pmatrix}.
\label{11}
\end{equation}

The standard procedure of the linear-response theory implies representation of solution of the equation $\imath \dot \rho=[H,\rho]$ for the atomic density matrix $\rho$ in the form $\rho=\rho_0+\rho_1+O(H_E^2)$ {(here $\rho_0$ is unperturbed atomic density matrix satisfying the equation
 $\imath\dot \rho_0=[H_0,\rho_0]$ and $\rho_1$ 
  is linear in the probe beam field  correction  satisfying the equation} 
  $\imath\dot\rho_1=[H_0,\rho_1]+[H_E,\rho_0]$),
 and computation of the quantities $\langle d_i\rangle$ as $\langle d_i\rangle = \mathop{\rm Sp}\rho_1d_i$, where $i=x,z$. It leads to the expression $\langle d_i\rangle =\alpha_{ik}A_{0k}({\bf R})$, in which the susceptibility tensor $\alpha$ has the following elements
\begin{equation}
\alpha_{ik}\equiv {d^2\over \hbar }\sum_{MM'M''}{\langle 1M|\rho_0|1M'\rangle \langle M'|J_k|M''\rangle \langle M''|J_i|M\rangle\over \Delta\omega+\imath\delta + \omega_{2L}M''-\omega_{1L}M' }, \qquad i,k=x,z.
\label{12}
\end{equation}

Here, $\Delta\omega\equiv \Omega-\omega$ is the optical detuning, $\imath\delta$ denotes the homogeneous broadening, $\langle M|J_k|M'\rangle$ are the matrix elements of the operator of $k$-th projection of the angular momentum $F$~\cite{Landau}, and the summation over $M,M'$, and $ M''$ is performed over $2F+1$ states of the ground-state multiplet. 
As has been noted above, the tensor $\alpha$ depends on time (through the matrix elements $\langle 1M|\rho_0|1M'\rangle $, see Eq.~\eqref{9}), with characteristic frequencies of this dependence corresponding to spectral features of the noise spectra observed in the SNS. 

{Since the quantity $\langle M|J_z|M'\rangle =\delta_{MM'}M$ and $\langle M|J_x|M'\rangle$ is nonzero only for $|M-M'|= 1$, it follows from Eq.~\eqref{12} that the only frequencies $\omega_{1L}|M-M'|$ at which oscillations of the tensor $\alpha$ may occur are:}
$\omega_{1L}$ (the elements $\alpha_{xz},\alpha_{zx}$ $\sim \langle M'|J_x|M''\rangle\langle M''|J_z|M\rangle$), 
$2\omega_{1L}$ (the elements $\alpha_{xx}$ $\sim \langle M'|J_x|M''\rangle\langle M''|J_x|M\rangle$) and 0 (the elements $\alpha_{zz},\alpha_{xx}$ $\sim \langle M'|J_x|M''\rangle\langle M''|J_x|M\rangle$, $\langle M|J_z|M''\rangle\langle M''|J_z|M\rangle) $). 

It is seen from Eq.~\eqref{8} for the complex polarimetric signal $\delta u$ that the components of this signal at the frequency $2\omega_L$ behave as $\sim \sin 2\theta$ and vanish when the probe beam polarization is parallel or perpendicular to the magnetic field ($\theta=0,\pi/2$), as it has been observed in experiment \cite{fomin-spin-align-noise2020}.

\section{Calculation of the polarimetric signal} 

Note that under assumption that $\omega_{1L}\ll\delta$ the dependence of the denominator in Eq.~\eqref{12} on the numbers $M''$ and $M'$ may be neglected. 
Then we obtain the following expression for the matrix of the atomic susceptibility $\alpha$ 
\begin{equation}
\alpha_{ik} ={d^2\hbox { Sp }\rho_0 J_kJ_i \over \hbar [\Delta\omega +\imath \delta]}
={d^2\hbox { Sp }\rho_0 [\{J_kJ_i\} +\imath\varepsilon_{kil}J_l]\over 2\hbar [\Delta\omega +\imath \delta]}, 
\label{13}
\end{equation} 
in which we distinguished symmetric ($\sim \{J_kJ_i\}\equiv J_kJ_i+J_iJ_k$) and antisymmetric (gyrotropic, $\sim J_kJ_i-J_iJ_k=\imath \varepsilon_{kil}J_l$) parts (here, $\varepsilon_{kil}$ is the Levi-Civita tensor).
After such a simplification, the expression for the susceptibility $\alpha$ acquires the form of a quantum mean value of the tensor observable with the operator $\sim J_kJ_i$ in the state with the density matrix $\rho_0$. 
The appropriate superpositional wavefunction $\Psi$ (it contains only the components related to the ground multiplet) satisfies the Schr\"odinger equation $\imath \dot \Psi=H_0\Psi = \omega_{1L}J_z\Psi$ and is defined by the formula: $\Psi(t)=\exp({-\imath \omega_{1L}J_zt})\Psi(0)$. 
Since the operator $\exp({-\imath \omega_{1L}J_zt})$ is the operator of rotation by the angle $\omega_{1L}t$ around the $z$~axis~\cite{Landau}, the function $\Psi(t)$ represents the function $\Psi(0)$, rotating around the magnetic field with the angular frequency $\omega_{1L}$. 
This rotation is accompanied by `rotation' of the tensor $\alpha_{ik}\sim \mathop{\rm Sp} \rho_0 J_kJ_i= \langle \Psi |J_kJ_i|\Psi\rangle$ \cite{f13}, and the noise signal detected in our experiments can be understood as a result of scattering of the probe beam by a quasi-point anisotropic system rotating with the Larmor frequency $\omega_{1L}$ around the magnetic field. 
If we substitute Eq.~\eqref{13} into \eqref{8}, \textit{}we obtain for the complex polarimetric signal $\delta u$ the following expression
\begin{equation}
\delta u= {\pi kd^2\over \hbar }|A_0({\bf R})|^2 f(t),
\label{14}
\end{equation}
where
\begin{equation*}
f(t) \equiv {\mathop{\rm Sp}\rho_0\bigg [ (J_x^2-J_z^2)\sin 2\theta +(J_zJ_x+J_xJ_z)\cos 2\theta +\imath J_y 
\bigg ]\over \Delta\omega+\imath \delta}\equiv f_e(t)+\imath f_r(t)
\end{equation*}

Physical meaning of different contributions in this formula can be determined by considering behavior of the function $f(t)$ at large detunings $\Delta\omega\gg\delta$. 
It can be seen that the first two terms in Eq.~\eqref{14} describe fluctuations of symmetric part of the tensor $\alpha$ (fluctuations of \emph{alignment}) and, being real (at $\Delta\omega\gg\delta$), can be observed only in the regime of detection of ellipticity (see Eq. (\ref{8})). 

Since the matrix elements $\langle M|J_x^2-J_z^2|M'\rangle$ are nonzero only at $|M-M'|=0$ and $|M-M'|=2$, the contribution $\sim (J_x^2-J_z^2)\sin 2\theta$ gives rise to peaks in the spectra of ellipticity noise at zeroth and double Larmor frequencies (because the elements $\langle M|J_x^2-J_z^2|M'\rangle$ enter the expression for the polarimetric signal Eq.\ (\ref{14}) together with the elements $\langle M'|\rho_0|M\rangle$ of the density matrix whose time behaviour is determined by Eq.\ (\ref{9})).
  In a similar way, one can make sure that the contribution $\sim (J_zJ_x+J_xJ_z)\cos 2\theta$ gives rise to a peak at the frequency $\omega_{1L}$. 

The isotropic term in brackets $\sim \imath \mathop{\rm Sp} \rho_0 J_y$ describes fluctuations of gyrotropy of the atomic system and, being pure imaginary, is revealed only in the FR noise. 
Since the matrix elements $\langle M|J_y|M'\rangle$ are nonzero only at $|M-M'|=1$, this term provides a feature in the Faraday-rotation noise spectrum only at the frequency $\omega_{1L}$.

The above consideration was related to the case of Voigt geometry. Similar results can be obtained for the Faraday geometry. In this case, the expression for the complex polarimetric signal $ \delta u $ differs from Eq.\ (\ref {14}) by the permutation of the operators $ J_z \rightarrow J_y $ and $ J_y \rightarrow J_z $ (leaving the same expression Eq.\ (\ref {9}) for the density matrix $ \rho_0 $). An analysis similar to the above shows that the polarimetric noise signal recorded in Faraday geometry will have spectral features only at zero frequency and at a frequency of $ 2 \omega_{1L}$.
 
The rigorous calculation of the noise power spectrum ${\cal N}(\nu)=\int \langle f(0)f(t)\rangle \exp({-\imath\nu t})dt$ observed in SNS experiments requires the calculation of the correlation function $\langle f (0)f (t)\rangle $. 
This calculation is somewhat cumbersome, so, in section \ref{sec:Dis}, we present the results of such calculation with no details. 
The quantum-mechanical correlation functions of the operators entering Eq.\ (\ref{14}) were calculated in \cite{fomin-spin-align-noise2020}.
 
\section{\label{sec:Dis} Discussion}
The above simplified consideration shows that observation of spectral feature at the frequency $2\omega_{1L}$ is possible only in the ellipticity noise spectrum. 
Note that experiments \cite{fomin-spin-align-noise2020} mainly support this conclusion.
A consistent calculation shows, however, that when the homogeneous width of the line $\delta$ is getting much smaller than the Doppler broadening, the difference between the noise spectra of ellipticity and FR (in terms of the peak at the double Larmor frequency) becomes not so dramatic. 
It can be briefly explained as follows. 
Consider, e.g., the ellipticity noise spectrum, which is determined by the Fourier-image of the correlation function $\langle \delta u_e(t)\delta u_e(0)\rangle\sim \langle |A_0({\bf R}(t))|^2|A_0({\bf R}(0))|^2 f_e(t)f_e(0)\rangle$ [see Eq.~\eqref{14}]. 
Calculation of correlator of the signal $f_e(t)$ \eqref{14} leads to the following expression:
	\begin{equation}
	\begin{gathered}
	\thinmuskip=0mu
    \medmuskip=0mu
    \thickmuskip=1mu
	\langle f_e(t)f_e(0)\rangle\sim  {5 a^2} \cos [\omega_{1L}t] +
	{d^2}\bigg [F(F+1)-{3\over 4}\bigg ]\bigg [ 4\cos^22\theta \cos [\omega_{1L}t]+ (3+\cos [2\omega_{1L}t]) \sin^2 2\theta \bigg ],\\
	d+\imath a\equiv 1/[\Delta \omega+kv_y +\imath \delta].
	\end{gathered}
	\label{15}
	\end{equation}
	
A similar expression was obtained in the theoretical section of work \cite{fomin-spin-align-noise2020} by solving the equations of motion for the correlation functions.
Here, we omitted nonessential factors and accounted for the Doppler shift $kv_y$ ($v_y$ is the projection of the atomic speed upon the probe beam direction). 
The expression for the correlator of the FR signal $\langle f_r(t)f_r(0)\rangle $ differs from Eq.~\eqref{15} by the substitutions 
$a\rightarrow d$ and $ d\rightarrow a$. 
Despite the fact that frequency dependencies of the correlators $\langle f_e(t)f_e(0)\rangle$ and $\langle f_r(t)f_r(0)\rangle $ are different, for both of them it has the form of a sharp feature with the width $\sim\delta$. 
For this reason, upon Maxwellian averaging of the Doppler shift $kv_y$, with the width $\sim kv_T\gg \delta$ (here, $v_T^2$ is the mean-square thermal velocity), the above difference (for $\Delta\omega \sim kv_T$) will be of no importance. 
For the Gaussian probe beam, calculation leads to the following expression for the correlation function observed in the SNS (nonessential numerical factors are omitted):
	\begin{equation}
	\begin{split}
	&K(t)\sim {\sigma \hskip1mm W^2 \rho_c k^2 d^4\hskip1mm
		\over v_T^2 \delta } 
	\exp \bigg [-{\Delta\omega^2 \over k^2v_T^2} \bigg ] {\exp({-|t|/T_2}) \over \sqrt{t^2+t_T^2}} \times \\
	&\times \bigg \{ 5 \cos [\omega_{1L}t] 
	+\bigg [F(F+1)-{3\over 4}\bigg ]\bigg [ 4\cos^22\theta \hskip1mm\cos [\omega_{1L}t]+ (3+\cos [2\omega_{1L}t])\hskip1mm
	\sin^2 2\theta \bigg ]\bigg \}.
	\end{split}
	\label{16}
	\end{equation}
Here, along with the quantities introduced above, we use: $\sigma$ -- atomic density, $W$ -- the probe beam power, $\rho_c$ -- the beam radius in its waist, $t_T\equiv \rho_c/ v_T$ -- the time of flight, and $T_2$ -- spin relaxation time. 
The polarization noise power ${\cal N}(\nu)$ is defined as ${\cal N}(\nu)=\int K(t) \exp({\imath \nu t}) dt$. 
As seen from Eq.~\eqref{16}, at $T_2,t_T\gg \omega_{1L}^{-1}$, the spectrum of the polarization noise power of the atomic vapor always reveals peaks at $\nu =0,\omega_{1L},$ and $ 2\omega_{1L}$. 
Angular dependence of amplitudes of these peaks at $F>1/2$ is controlled by the last term in brackets and $\sim \cos^2 2\theta$ for the peak at $\nu=\omega_{1L}$ and $\sim \sin^2 2\theta$ for the peak at $\nu=2\omega_{1L}$. 

\begin{figure*}[t]
\centering
	\subfigure{\label{fig3_a}}
	\subfigure{\label{fig3_b}}
	\subfigure{\label{fig3_c}}
	\includegraphics[width=\textwidth,clip]{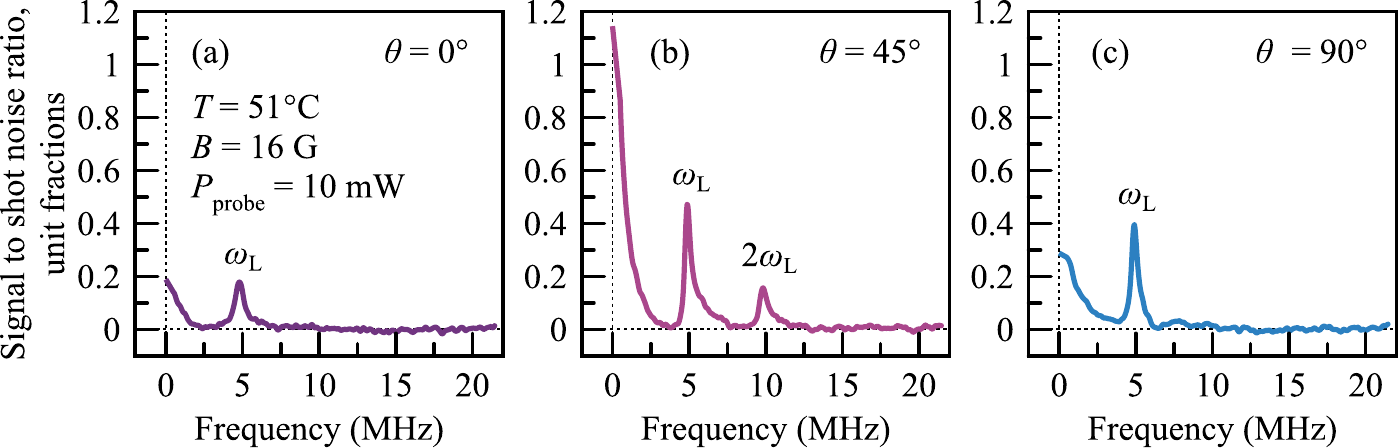}
	\caption{Ellipticity noise spectra detected at angles $\theta = 0^\circ$ (a), $45^\circ$ (b) and $90^\circ$ (c) between the probe beam polarization plane and magnetic field direction. The second harmonic of the Larmor frequency is well pronounced at $\theta=45^\circ$ and vanishes at $\theta = 0$ and $90^\circ$.}
	\label{fig3}
\end{figure*}

Dependence Eq.\ (\ref{16}) of the ellipticity noise spectra on azimuth $\theta$ of the polarization plane of the probe beam qualitatively agrees with the experimental data (see Fig.~\ref{fig3}, which represents data obtained in our work~\cite{fomin-spin-align-noise2020}), in particular, the amplitude of the peak at the frequency $2\omega_{1L}$ reaches maximum at $\theta=\pi/4$ and vanishes at $\theta=0$ and $\pi/2$. 

{The following remark should be made regarding Fig.~\ref{fig3} which shows that at $\theta=\pi/4$, the amplitude of the peak at the Larmor frequency $\omega_{1L}$ exceeds the amplitude of the maximum at twice the Larmor frequency $2\omega_{1L}$. This contradicts formula Eq.\ (\ref{16}), according to which, for $F>1/2$, the amplitude of the peak at the Larmor frequency $\omega_{1L}$ should be less than that of the peak at doubled Larmor frequency. We attribute this discrepance to nonlinear effects that took place in experiments \cite{fomin-spin-align-noise2020} which were carried out under conditions of resonant probing of an atomic system with a laser beam. These effects require special consideration, which we plan to present in our future publications.}

\section{\label{sec:conclusion}Conclusion}

The signals observed in spin noise spectroscopy (SNS) are analyzed from the view point of scattering theory. Expressions connecting SNS signals with fluctuations of the optical susceptibility tensor are obtained.
		This tensor is calculated in the framework of the simplest atomic model, and it is shown that the fluctuations of its gyrotropic (antisymmetric) part are proportional to the magnetization fluctuations and determine the Faraday rotation noise at Larmor frequency observed in typical SNS experiments. However, the polarization noise of the scattered optical field is not exhausted by the noise of its Faraday rotation. It is shown, in this paper, that fluctuations of the symmetric part of the optical susceptibility tensor (the alignment noise) give rise to the ellipticity noise of the scattered field spectrally localized at the double Larmor frequency. 
		Using the developed ideas, we interpret the experiment \cite{fomin-spin-align-noise2020} where the ellipticity noise at a double Larmor frequency was observed.
		Our results are in agreement with those obtained in \cite{fomin-spin-align-noise2020} by means of symmetry analysis of quantum correlation functions of the system under study.
		 
Note, in conclusion, that atomic systems showing strong and spectrally narrow signals of spin noise can be considered as highly convenient model objects of the spin noise spectroscopy. 
At the same time, theoretical description of the spin noise formation, in these systems, especially under conditions of strong resonant probing, proves to be rather complicated due to complex energy structure of the dynamically broadened atomic states. 
The proposed theoretical model provides the the most consistent basis for accurate description of the effects of nonlinear spectroscopy of spin noise (in the presence of optical pumping, optical Stark effect, optically induced spin dephasing, etc.).

{The mechanism of the formation of noise polarimetric signals considered in our article can possibly be applied not only to gas, but also to solid-state systems, including nano-systems such as quantum wells in microcavities, quantum wires, etc., noise signals in which were calculated using the "magnetic language" \cite{Sherman1,Sherman2,Sherman3}. In this case, a consistent calculation of optical scattering can be important for describing diffraction effects and effects associated with the internal motion of excitations in such systems. A more detailed discussion of these effects is beyond the scope of this article. }
Further development of this approach can make important contribution to physics of the light-matter interaction. 

\section*{Acknowledgements}
The authors are grateful to M.~M.~Glazov for useful discussions.

The work was supported by the RFBR Grant No. 19-52-12054 which is highly appreciated. The authors acknowledge Saint-Petersburg State University for the research Grant No. 51125686. 
Experimental part of the work was performed by A.\,A.\,F. and M.\,Y.\,P. under support of the Russian Science Foundation (Grant No. 18-72-00078) and partially using equipment of the SPbU Resource Center ``Nanophotonics''.

\section*{Disclosures}
The authors declare no conflicts of interest.

\section*{Appendix}
\begin{figure}[b]
	\begin{center}
		\includegraphics[width=8cm]{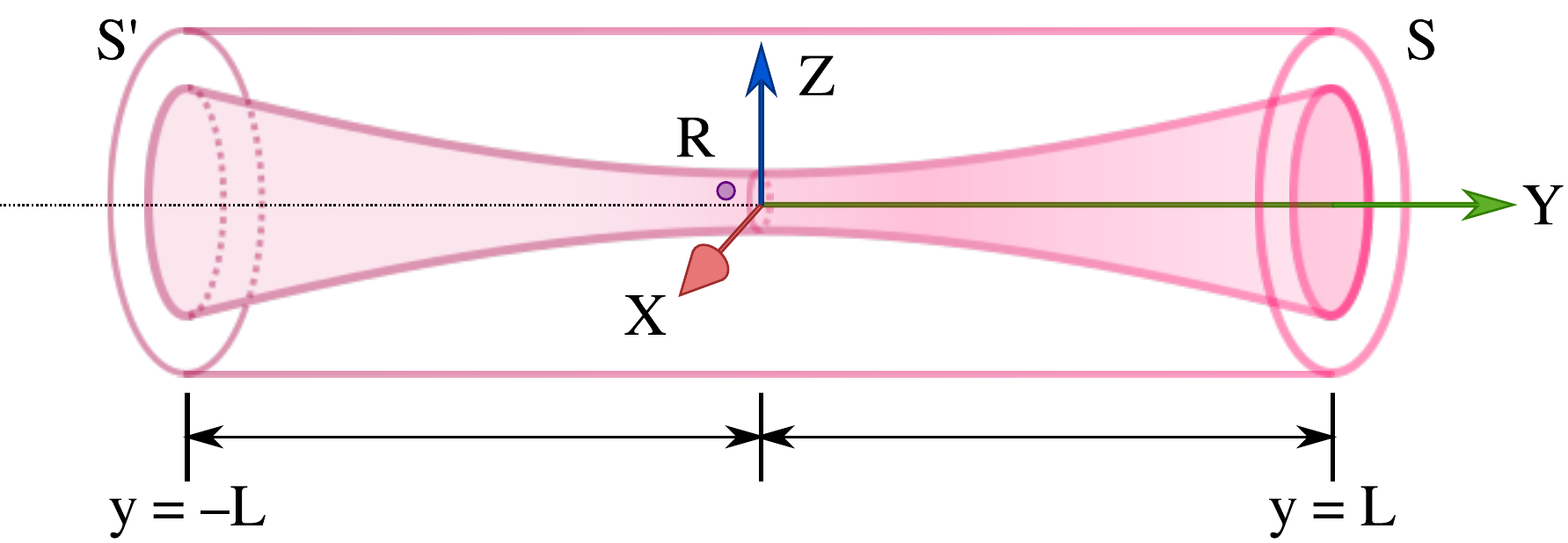}
		\caption{See explanations in the text.}
		\label{Kirh}
	\end{center}
\end{figure}

We present here calculation of the function
$ \Phi_i ^ - $ [Eq. (\ref {5})]. Let us introduce the auxiliary functions: 
${\bf F} ({\bf r'})\equiv \int_S dxdz\hskip1mm {\bf E}_0({\bf r})\Gamma({\bf r-r'})\sim \exp({-\imath\omega t})$
and ${\bf P} ({\bf r'})\equiv \int_S dxdz\hskip1mm {\bf E}_0^\ast({\bf r})\Gamma({\bf r-r'})\sim \exp({\imath\omega t})$ (do not confuse with polarization denoted by the same letter in the main text). 
Then function $ \Phi_ \alpha ({\bf r '})$ Eq. (\ref{4}) can be represented as $ \Phi_ \alpha ({\bf r '}) = - 2 \pi [F_ \alpha ({\bf r'}) + P_ \alpha ({\bf r '})] $
and therefore $ \Phi ^ -_ \alpha ({\bf r '}) e ^ {\imath \omega t} = - 2 \pi P_ \alpha ({\bf r'}) $. To calculate the integral in the above expression
for $ P_ \alpha ({\bf r '})$, we 
 note that the field $ {\bf E} _0 ^ \ast ({\bf r}) $ satisfies the Helmholtz equation, 
 $\Delta {\bf E}_0^\ast+k^2{\bf E}_0^\ast=0$
 and, therefore,
 the Kirchhoff formula \cite {Born-Wolf} can be used:

\begin{equation}
{\bf E}_0^\ast ({\bf R})=\int_ {\partial V} \bigg [{\bf E}_0^\ast({\bf r}){\partial \Gamma({\bf r-R})\over \partial n} 
-\Gamma({\bf r-R}){\partial {\bf E}_0^\ast({\bf r})\over \partial n}\bigg ]d S
\label{30p}
\end{equation}
Here, $ \Gamma({\bf r})=-{\exp({\imath kr})/4\pi r}$ is the Green's function of Helmholtz operator, $ \partial V $ is an arbitrary closed surface surrounding the point $ \bf R $, and the symbol $ \partial / \partial n $ means the normal derivative to the surface $ \partial V $ \cite{f14}.
For definiteness, we assume that the photosensitive surface of photodetectors $S$ is larger than the 'cross section'
of the probe Gaussian beam, and the photodetectors "intercept" almost the entire flow of its energy
(recall that $ S $ lies in the plane $ y = L \gg | {\bf R} | $). In addition, we assume that the size of the surface $S$ 
(we denote this size by the symbol $ \sqrt S $) is much smaller than the distance $ L $ from the scattering particle to the photodetectors: $ \sqrt S \ll L $.
Now we choose the surface $ \partial V $ in the form of a cylinder with its axis parallel to the axis $ y $ and leaning on the right
to the surface $ S, (y = L) $ and on the left - to the surface $ S '$ symmetric with respect to the origin, ($ y = -L $)
(see Fig. \ref {Kirh}).
Since the probe beam is located inside the cylinder constructed in this way, the integral in the Kirchhoff formula (\ref {30p}) can be calculated only over the surfaces $ S $ and $ S '$, because on the side surface of the cylinder
the field of the probe beam is negligible. Both surfaces are perpendicular to the $ y $ axis
and have oppositely directed normal vectors. Therefore, on the surface $ S $ the normal derivative is calculated as
$ \partial / \partial n \rightarrow \partial / \partial y $, and as
$ \partial / \partial n \rightarrow - \partial / \partial y $ on the surface $ S '$.
We calculate the derivative of the Green's function $ \partial \Gamma ({\bf r-R}) / \partial n $ on the surface $ S $ (that is, for $ y = L $):
\begin{equation}
{\partial \Gamma ({\bf r-R})\over \partial n}\bigg |_S={\partial \Gamma({\bf r-R})\over \partial y}= {\partial \Gamma({\bf r-R})\over \partial |{\bf r-R}|} 
{\partial |{\bf r-R}|\over \partial y}=
\end{equation}
$$
={\partial \Gamma({\bf r-R})\over \partial |{\bf r-R}|} 
{y-R_y\over |{\bf r-R}|}=
$$
$$
=\Gamma({\bf r-R})\bigg [\imath k-\overbrace {{1\over |{\bf r-R}|}}^{\approx L^{-1}\ll k}\bigg ]
\overbrace {{L-R_y\over |{\bf r-R}|}}^{\approx 1}\approx
\imath k \Gamma({\bf r-R})
$$
The approximations used are justified, since
$ R, \sqrt S \ll L $ ($ R \equiv | {\bf R} | $) and therefore $ | L-R_y |, | {\bf r-R} | \sim L $.
This allows us to neglect the second term in the expression in square brackets, since
$ | {\bf r-R} | ^ {- 1} \sim L ^ {- 1} \ll k = 2 \pi / \lambda $. In addition, the factor $ [y-R_y] / | {\bf r-R} | $ for
$ r, y \gg R $ is practically equal to 1 on the surface $ S $.
The derivative of the Green's function $ \partial \Gamma ({\bf r-R}) / \partial n $
on the surface $ S '$ (i.e. for $ y = -L $):
\begin{equation}
{\partial \Gamma ({\bf r-R})\over \partial n}\bigg |_{S'}=-
{\partial \Gamma({\bf r-R})\over \partial y}=
\end{equation}
$$
=- {\partial \Gamma({\bf r-R})\over \partial |{\bf r-R}|} 
{\partial |{\bf r-R}|\over \partial y}=-{\partial \Gamma({\bf r-R})\over \partial |{\bf r-R}|} 
{y-R_y\over |{\bf r-R}|}=
$$
$$
=-\Gamma({\bf r-R})\bigg [\imath k-\overbrace {{1\over |{\bf r-R}|}}^{\approx L^{-1}\ll k}\bigg ]
\overbrace {{-L-R_y\over |{\bf r-R}|}}^{\approx -1}\approx
\imath k \Gamma({\bf r-R})
$$
Thus, we obtain that on both surfaces $ S $ and $ S '$ the following relation holds
\begin{equation}
{\partial \Gamma({\bf r-R})\over \partial n}=\imath k \Gamma({\bf r-R})
\end{equation}
Let us turn now to calculation of the
derivative $ {\partial {\bf E} _0 ^ \ast ({\bf r}) / \partial n}$
 entering Eq. (\ref {30p}).
First, we calculate it on the surface $ S $, where
$ \partial / \partial n = \partial / \partial y $. 
We now take into account that the probe beam field ${\bf E}_0$ is close to that of the plane wave $\sim \exp({\imath k y})$ propagating along $y$-axis. For this reason, we can write the following expression for the considered derivative

\begin{equation}
{\partial {\bf E}_0^\ast({\bf r})\over \partial n}\bigg |_S={\partial {\bf E}_0^\ast({\bf r})\over \partial y}=
-\imath k {\bf E}_0^\ast ({\bf r})
\end{equation} 
A similar calculation of the derivative
$ {\partial {\bf E} _0 ^ \ast ({\bf r}) / \partial n} $ on the surface $ S '$ leads to the expression
\begin{equation}
{\partial {\bf E}_0^\ast({\bf r})\over \partial n}\bigg |_{S'}=-{\partial {\bf E}_0^\ast({\bf r})\over \partial y}=
\imath k {\bf E}_0^\ast({\bf r})
\end{equation} 

Substituting the obtained relations into formula (\ref {30p}), we obtain
\begin{equation}
{\bf E}_0^\ast({\bf R})=\int_ {S} \bigg [{\bf E}_0^\ast({\bf r}){\partial \Gamma({\bf r-R})\over \partial n} 
-\Gamma({\bf r-R}){\partial {\bf E}_0^\ast({\bf r})\over \partial n}\bigg ]d S+
\label{36p}
\end{equation}
$$
+\overbrace {\int_ {S'} \bigg [{\bf E}_0^\ast({\bf r}){\partial \Gamma({\bf r-R})\over \partial n} 
	-\Gamma({\bf r-R}){\partial {\bf E}_0^\ast({\bf r})\over \partial n}\bigg ]d S}^{=0}=
$$
$$
=2\imath k\int_ {S} {\bf E}_0^\ast({\bf r}) \Gamma({\bf r-R}) \hskip1mm dS=2\imath k \hskip1mm {\bf P}({\bf R}) 
$$
Taking into account that $ \Phi ^ -_ \alpha ({\bf r '}) e ^ {\imath \omega t} = - 2 \pi P_ \alpha ({\bf r'}) $, we have
$
E_{0\alpha }^\ast({\bf R})= -{\imath k\over \pi} \Phi_\alpha^- ({\bf R}) \exp({\imath\omega t}).
$ 
Bearing in mind that ${\bf E}_0=(A_{0x},0,A_{0z})\exp({-\imath\omega t})$,
 we come to the result (\ref {5}). 

\bibliography{raman}

\begin{thebibliography}{10}
\bibitem{for}
Upon receipt of this formula, generally speaking, four terms arise. However,
  proper replacement of the variable during integration over time and doubling
  of the result allows us to obtain Eq.\ (\ref{2}).
\bibitem{f0}
The function $\Phi({\bf r})$ has the sense of the field created by the source
  distributed over the surface of the detector with the density ${\bf {\cal
  E}}_0(x,y,L)$. For this reason, the field $\Phi({\bf r})$ is similar to that
  of the main beam ${\bf {\cal E}}_0({\bf r})$. Appropriate formula is derived
  in Appendix and in \cite{kozlov-heterodyne-detect-scatt-tomography2018} and
  is given below (Eq.\ (\ref{5})).
\bibitem{fbs}
Rotation of the half-wave plate in the polarimetric detector (see
  Fig.\ref{fig1}) by any angle $\xi$ is equivalent to changing $\phi$ by
  $\phi+2\xi$. Therefore, the polarimetric detector can be balanced by rotating
  the half-wave plate (see text below). For the introduced above coordinate
  system with $z$-axis parallel to main direction of beamsplitter, we can set
  $\phi=2\xi$.
\bibitem{f12}
In experiment \cite{fomin-spin-align-noise2020}, the hyperfine component of the
  cesium $D2$ line is associated with transitions from the multiplet $F=4$ to
  the multiplets $F=3,4,5$, but for the semiquantitative interpretation of the
  observed effects this simplification seems acceptable.
\bibitem{f13}
Mentioned rotation is the rotation of the principal axes of the symmetric part
  of the tensor $\alpha$ and the rotation of the gyration vector corresponding
  to its antisymmetric part.
\end{thebibliography}

\end{document}